\documentclass{article}
\usepackage{amssymb,amsmath}
 \allowdisplaybreaks
\begin{document}

\title{FIRST-CLASS APPROACHES TO MASSIVE $2$-FORMS}

\author{E. M. Cioroianu\thanks{ e-mail address:
manache@central.ucv.ro}, S. C. Sararu\thanks{ e-mail
address: scsararu@central.ucv.ro}, O. Balus\\
Faculty of Physics, University of Craiova,\\ 13 Al. I. Cuza Street
Craiova, 200585, Romania}
\date{}
\maketitle

\begin{abstract}
Massive $2$-forms are analyzed from the point of view of the
Hamiltonian quantization using the gauge-unfixing approach and respectively
the Batalin--Fradkin method. Both methods finally output the manifestly
Lorentz covariant path integral for $1$- and $2$-forms with St\"{u}ckelberg
coupling.
\end{abstract}


\section{Introduction}

Models with $p$-form gauge fields (antisymmetric tensor fields of various
orders) play an important role in string and superstring theory,
supergravity and the gauge theory of gravity \cite{str}--\cite{pbrane}.
Antisymmetric tensor fields of various orders are included within the
supergravity multiplets of many supergravity theories \cite{sgr}--\cite%
{sgrav0}, especially in 10 or 11 dimensions. Moreover, $p$-forms have a
special place in the theory of $p$-branes \cite{pbrane}, where $(p+1)$-forms
couple naturally to $p$-branes. In fact, it is known that the configuration
space for closed $p$-branes is nothing but the space of all closed $p$%
-manifolds embedded in space-time, in which background rank-$(p+1)$
antisymmetric tensor fields should be analyzed in connection with their
geometric aspects. Interacting $p$-form gauge theories have been analyzed
from the redundant Hamiltonian BRST point of view in \cite{marcpform}, where
the ghost and auxiliary field structures required by the antifield BRST
formalism are derived. Finally, it is worth to notice that a $U(1)$ gauge theory
defined in the configuration space for closed $p$-branes yields the gauge
theory of a massless rank-$(p+1)$ antisymmetric tensor field plus the St\"{u}%
ckelberg formalism for a massive vector field. From the point of view of
Hamiltonian second-class constrained systems, the St\"{u}ckelberg formalism
has been largely used at the quantization of massive vector field following
various schemes \cite{stuck1}--\cite{stuckn}.

The main aim of this paper is to quantize massive 2-forms using two
different methods: gauge-unfixing \cite{jap}--\cite{mitra} and
Batalin--Fradkin \cite{bfgen}--\cite{battyut}. The first approach (gauge
unfixing method) \cite{jap}--\cite{mitra} relies on separating the
second-class constraints into two subsets, one of them being first-class and
the other providing some canonical gauge conditions for the first-class
subset. Starting from the canonical Hamiltonian of the original second-class
system, one constructs a first-class Hamiltonian with respect to the
first-class subset through an operator that projects any smooth function
defined on the phase-space into a function that is in strong involution with
the first-class subset. A systematic BRST treatment of the gauge-unfixed method has been realized in Refs.~\cite{gubrst1} and \cite{gubrst2}. The second approach (Batalin--Fradkin method) \cite%
{bfgen}--\cite{battyut} relies on enlarging the original phase-space and
constructing a first-class constraint set and a first-class Hamiltonian,
with the property that they coincide with the original second-class
constraints and respectively with the starting canonical Hamiltonian if one
sets all the extravariables equal to zero.

This paper is organized in four sections. In Section \ref{sect2} we start
from a bosonic second-class constrained system and briefly expose the above
mentioned methods of constructing first-class systems equivalent with the
original theory. In Section \ref{sect3} we apply both methods to massive $2$-forms and meanwhile obtain the path integrals corresponding to
the first-class systems associated with this model. After integrating out
the auxiliary fields and performing some field redefinitions, we discover
nothing but the manifestly Lorentz covariant path integrals corresponding to
the Lagrangian formulation of the first-class systems, which reduce to the
Lagrangian path integral for St\"{u}ckelberg-coupled $1$- and $2$-forms.
Section \ref{sect4} ends the paper with the main conclusions.

\section{First-Class Approaches to Second-Class Constrained Systems\label%
{sect2}}

The starting point is a bosonic dynamic system with the phase-space locally
parameterized by $n$ canonical pairs $z^{a}=\left( q^{i},\ p_{i}\right) $,
endowed with the canonical Hamiltonian $H_{c}$, and subject to the purely
second-class constraints%
\begin{equation}
\chi _{\alpha _{0}}\left( z^{a}\right) \approx 0,\qquad \alpha _{0}=%
\overline{1,2M_{0}},  \label{ec3}
\end{equation}%
where \textquotedblleft $\approx $\textquotedblright\ represents the weak
equality symbol. The idea is to associate a first-class system with the
original second-class theory that satisfies the following requirements: its
number of physical degrees of freedom coincides with that of the original
second-class theory, the algebras of classical observables are isomorphic,
the first-class Hamiltonian (governing the dynamics of the first-class
system) restricted to the constraint surface (\ref{ec3}) reduces to the
original canonical Hamiltonian $H_{c}$. The construction of such a
first-class system, equivalent to a given, second-class one, can proceed in
several ways. As announced in the introduction, we chose two of them. One is
based on interpreting the second-class constraint set as stemming from the
gauge-fixing of a first-class constraint set \cite{jap}--\cite{mitra} and
the other on enlarging (in an appropriate manner) the original phase-space
and constructing a first-class constraint sets that reduces to (\ref{ec3})
in the zero limit of all extravariables \cite{bfgen}--\cite{battyut}.

\subsection{Gauge unfixing (GU) method\label{GUmeth}}

Assume that one can split the second-class constraint set (\ref{ec3}) into
two subsets with equal numbers of independent constraint functions%
\begin{equation}
\chi _{\alpha _{0}}\left( z^{a}\right) \equiv \left( G_{\bar{\alpha}%
_{0}}\left( z^{a}\right) ,C^{\bar{\beta}_{0}}\left( z^{a}\right) \right)
\approx 0,\qquad \bar{\alpha}_{0},\bar{\beta}_{0}=\overline{1,M_{0}},
\label{ec4}
\end{equation}%
such that
\begin{equation}
\left[ G_{\bar{\alpha}_{0}},G_{\bar{\beta}_{0}}\right] =D_{\bar{\alpha}_{0}%
\bar{\beta}_{0}}^{\bar{\gamma}_{0}}G_{\bar{\gamma}_{0}},  \label{ec5}
\end{equation}%
where $D_{\bar{\alpha}_{0}\bar{\beta}_{0}}^{\bar{\gamma}_{0}}$ may in
principle be functions of $z^{a}$. Hereafter the square brackets, $\left[ ,\right] $, denote the Poisson brackets on
the phase space of the theory. On the one hand, relations (\ref{ec5}%
) yield the subset
\begin{equation}
G_{\bar{\alpha}_{0}}\left( z^{a}\right) \approx 0  \label{ec6}
\end{equation}%
to be first-class. On the other hand, the second-class behaviour of the
overall constraint set ensures that
\begin{equation}
C^{\bar{\alpha}_{0}}\left( z^{a}\right) \approx 0  \label{ec7}
\end{equation}%
may be regarded as some gauge-fixing conditions for this first-class set. It
is possible to construct a first-class Hamiltonian with respect to (\ref{ec6}%
) with the help of an operator $\hat{X}$ \cite{vytgen}--\cite{vyt2} that
associates with every smooth function $F$ on the original phase-space an
application $\hat{X}F$, which is in strong involution with the functions $G_{%
\bar{\alpha}_{0}}$,%
\begin{gather}
\hat{X}F=F-C^{\bar{\alpha}_{0}}\left[ G_{\bar{\alpha}_{0}},F\right] +\frac{1%
}{2}C^{\bar{\alpha}_{0}}C^{\bar{\beta}_{0}}\left[ G_{\bar{\alpha}_{0}},\left[
G_{\bar{\beta}_{0}},F\right] \right] -\cdots ,  \label{ec8} \\
\left[ \hat{X}F,G_{\bar{\alpha}_{0}}\right] =0.  \label{ec8a}
\end{gather}%
If we denote by $\mathcal{S}_{O}$ and $\mathcal{S}_{GU}$ the original and
respectively the gauge-unfixed system, then they are classically equivalent
since they possess the same number of physical degrees of freedom%
\begin{equation}
\mathcal{N}_{O}=\frac{1}{2}\left( 2n-2M_{0}\right) =\mathcal{N}_{GU}
\label{ec9}
\end{equation}%
and, moreover, the corresponding algebras of classical observables are
isomorphic%
\begin{equation}
Phys\left( \mathcal{S}_{O}\right) =Phys\left( \mathcal{S}_{GU}\right) .\label{ec10}
\end{equation}%
Consequently, the two systems become also equivalent at the level of the
path integral quantization, which allows one to replace the Hamiltonian path
integral of the original second-class theory
\begin{equation}
Z_{O}=\int \mathcal{D}\left( z^{a},\lambda ^{\alpha _{0}}\right) \det \left( %
\left[ G_{\bar{\alpha}_{0}},C^{\bar{\beta}_{0}}\right] \right) \exp \left[
\mathrm{i}\int dt\left( \dot{q}^{i}p_{i}-H_{c}-\lambda ^{\alpha _{0}}\chi
_{\alpha _{0}}\right) \right]   \label{ec11}
\end{equation}%
with that of the gauge-unfixed first-class system%
\begin{eqnarray}
Z_{GU} &=&\int \mathcal{D}\left( z^{a},\lambda ^{\bar{\alpha}_{0}}\right)
\left( \prod\limits_{\bar{\alpha}_{0}} \delta\left( C^{\bar{\alpha}%
_{0}}\right) \right) \left( \det \left( \left[ G_{\bar{\alpha}_{0}},C^{\bar{%
\beta}_{0}}\right] \right) \right) \times   \notag \\
&&\times \exp \left[ \mathrm{i}\int dt\left( \dot{q}^{i}p_{i}-\hat{X}%
H_{c}-\lambda ^{\bar{\alpha}_{0}}G_{\bar{\alpha}_{0}}\right) \right] .
\label{ec12}
\end{eqnarray}%
In the above $\lambda ^{\alpha _{0}}$ denote the Lagrange multipliers
associated with the constraints (\ref{ec4}), while $\lambda ^{\bar{\alpha}%
_{0}}$ correspond to the first-class subset (\ref{ec6}). For concrete
models, the argument of the exponential from the path integral (\ref{ec12})
may contain other terms as well, such that the integration measure should be
accordingly modified \cite{marcel}.

\subsection{Batalin-Fradkin (BF) method\label{BFmeth}}

The BF approach \cite{bfgen}--\cite{battyut} to the problem of constructing
a first-class system equivalent to the starting second-class one (subject to
the second-class constraints (\ref{ec3})) relies on enlarging the original
phase-space with $2M$ ($M\geq M_{0}$) bosonic variables $\left( \zeta
^{\alpha }\right) _{\alpha =\overline{1,2M}}$ and on further extending the
Poisson bracket to the newly added variables through the relations%
\begin{equation}
\left[ \zeta ^{\alpha },z^{a}\right] =0,\qquad \left[ \zeta ^{\alpha },\zeta
^{\beta }\right] =\omega ^{\alpha \beta }.  \label{bf9}
\end{equation}%
In the aboveIn the above $\omega ^{\alpha \beta }$ are the elements of a quadratic,
antisymmetric and invertible matrix, independent of the extended phase-space variables. The
elements of its inverse will be denoted by $\omega _{\alpha \beta }$ in the
sequel. The next step is to construct a set of independent, smooth, real
functions defined on the extended phase-space, $\left( G_{A}\left( z,\zeta
\right) \right) _{A=\overline{1,M_{0}+M}}$, such that it reduces to the
original second-class constraint function set $\left( \chi _{\alpha
_{0}}\left( z\right) \right) _{\alpha _{0}=\overline{1,2M_{0}}}$ in the
limit of setting all the extravariables equal to zero%
\begin{eqnarray}
G_{\alpha _{0}}\left( z,0\right) &\equiv &\chi _{\alpha _{0}}\left( z\right)
,\qquad \alpha _{0}=\overline{1,2M_{0}},  \label{bf10a} \\
G_{\bar{A}}\left( z,0\right) &\equiv &0,\qquad \bar{A}=\overline{%
2M_{0}+1,M_{0}+M},  \label{bf10b}
\end{eqnarray}%
and, moreover, the functions $G_{A}$ are in (strong) involution
\begin{equation}
\left[ G_{A},G_{B}\right] =0,\qquad A,B=\overline{1,M_{0}+M}.  \label{bf11}
\end{equation}%
In the last step one generates a smooth, real function, defined on the
extended phase-space, $H_{BF}=H_{BF}\left( z,\zeta \right) $, with the
properties that $H_{BF}$ reduces to $H_{c}$ in the limit of setting all the
extravariables equal to zero%
\begin{equation}
H_{BF}\left( z,0\right) \equiv H_{c}\left( z\right)  \label{bf12}
\end{equation}%
and is in involution with the first-class constraint functions $%
\left( G_{A}\right) _{A=\overline{1,M_{0}+M}}$
\begin{equation}
\left[ H_{BF},G_{A}\right] =V_{A}^{\quad B}G_{B},\qquad A=\overline{1,M_{0}+M%
}.  \label{bf13}
\end{equation}%
The previous steps unravel a dynamic system subject to the first-class
constraints%
\begin{equation}
G_{A}\left( z,\zeta \right) \approx 0,\qquad A=\overline{1,M_{0}+M},
\label{bf14}
\end{equation}%
whose evolution is governed by the first-class Hamiltonian $%
H_{BF}=H_{BF}\left( z,\zeta \right) $. Denoting by $\mathcal{S}_{BF}$ the BF
first-class system, it follows that it is classically equivalent to the
original theory $\mathcal{S}_{O}$ since both of them display the same number
of physical degrees of freedom
\begin{equation}
\mathcal{N}_{O}=\frac{1}{2}\left( 2n-2M_{0}\right) =\frac{1}{2}\left[
2n+2M-2\left( M_{0}+M\right) \right] =\mathcal{N}_{BF}  \label{bf15}
\end{equation}%
and, in addition, the corresponding algebras of classical observables are
isomorphic%
\begin{equation}
Phys\left( \mathcal{S}_{O}\right) =Phys\left( \mathcal{S}_{BF}\right) .
\label{bf16}
\end{equation}%
In turn, the above isomorphism renders the two systems equivalent also at
the level of the path integral quantization and hence allows the replacement
of the Hamiltonian path integral for the original second-class theory with
that of the BF first-class system.

\section{The Model\label{sect3}}

We start from the Lagrangian action of massive $2$-forms in $D$
space-time dimensions ($D\geq 3$) \cite{marcpform},\cite{biz2form}%
\begin{equation}
S_{0}^{L}\left[ A_{\mu \nu }\right] =\int d^{D}x\left( -\frac{1}{12}F_{\mu
\nu \rho }F^{\mu \nu \rho }-\frac{m^{2}}{4}A_{\mu \nu }A^{\mu \nu }\right) ,
\label{act2fm}
\end{equation}%
with $m$ the mass of $A_{\mu \nu }$ and $F_{\mu \nu \rho }$ the field
strength of the $2$-form, defined in the standard manner as $F_{\mu
\nu \rho }=\partial _{\lbrack \mu }A_{\nu \rho ]}\equiv \partial _{\mu
}A_{\nu \rho }+\partial _{\nu }A_{\rho \mu }+\partial _{\rho }A_{\mu \nu }$.
Everywhere in this paper the notation $[\mu \nu \ldots \rho ]$ signifies
complete antisymmetry with respect to the (Lorentz) indices between
brackets, with the conventions that the minimum number of terms is always
used and the result is never divided by the number of terms. We work with
the Minkowski metric tensor of `mostly minus' signature $\sigma _{\mu \nu
}=\sigma ^{\mu \nu }=\mathrm{diag}\left( +-\ldots -\right) $. In the sequel
we denote by $\pi ^{\mu \nu }$ the canonical momenta respectively conjugated
with $A_{\mu \nu }$. For definiteness, we work with the non-vanishing
fundamental Poisson brackets%
\begin{eqnarray}
\left[ A_{0i}(x),\pi ^{0j}(y)\right] _{x^{0}=y^{0}} &=&\delta _{i}^{j}\delta
\left( \mathbf{x}-\mathbf{y}\right) ,  \label{PB1} \\
\left[ A_{ij}(x),\pi ^{kl}(y)\right] _{x^{0}=y^{0}} &=&\frac{1}{2}\delta
_{\lbrack i}^{k}\delta _{j]}^{l}\delta \left( \mathbf{x}-\mathbf{y}\right) .
\label{PB2}
\end{eqnarray}%
By performing the canonical analysis of this model \cite{dirac1}--\cite%
{dirac2}, there result the constraints
\begin{eqnarray}
\chi ^{\left( 1\right) i} &\equiv &\pi ^{0i}\approx 0,  \label{gei} \\
\chi _{i}^{\left( 2\right) } &\equiv &2\partial ^{j}\pi _{ji}-m^{2}A_{0i}\approx 0,
\label{cei}
\end{eqnarray}%
along with the canonical\ Hamiltonian
\begin{equation}
H_{c}(x^{0})=\int d^{D-1}x\left( -\pi _{ij}\pi ^{ij}+\frac{1}{12}%
F_{ijk}F^{ijk}+\frac{m^{2}}{4}A_{\mu \nu }A^{\mu \nu }-2A_{0i}\partial
_{j}\pi ^{ji}\right) .  \label{hcan}
\end{equation}%
The constraints (\ref{gei}) and (\ref{cei}) are second-class and irreducible
(see Ref.~\cite{marc}, Chapter 1, Subsection 1.3.4), with the matrix of the Poisson brackets among the constraint
functions expressed by%
\begin{equation}
\left( \left[ \chi _{\alpha _{0}}(x),\chi _{\beta _{0}}(y)\right]
_{x^{0}=y^{0}}\right) =\left(
\begin{array}{cc}
\mathbf{0} & m^{2}\delta _{j}^{i} \\
-m^{2}\delta _{i}^{j} & \mathbf{0}%
\end{array}%
\right) \delta \left( \mathbf{x}-\mathbf{y}\right) ,  \label{cab}
\end{equation}%
so the matrix (\ref{cab}) is invertible. The number of physical degrees of
freedom per space point is equal to $\mathcal{N}_{O}=\left( D-1\right)
\left( D-2\right) /2$.

\subsection{GU method}

According to the GU method exposed in subsection \ref{GUmeth}, one may
consider either of the constraints (\ref{gei}) or (\ref{cei}) as the
first-class constraint set and the remaining constraints ((\ref{cei}) or
respectively (\ref{gei})) as the corresponding canonical gauge conditions.
The first choice ((\ref{gei}) are first-class and (\ref{cei}) their
associated gauge-fixing conditions) yields a path integral that cannot be
written in a manifestly covariant form. This can be shown for instance along
a line similar to that employed in \cite{vyt3} with respect to the Proca
field, and therefore we will avoid this choice. Thus, we adhere to the
second choice and redefine the first-class constraints (\ref{cei}) as
\begin{equation}
G^{i}\equiv -\frac{1}{m^{2}}\left( 2\partial _{j}\pi
^{ji}-m^{2}A^{0i}\right) \approx 0.  \label{cein}
\end{equation}%
The first-class Hamiltonian with respect to (\ref{cein}) follows from
relation (\ref{ec8}), with $H_{c}$ expressed by (\ref{hcan}), and reads as
\begin{eqnarray}
&&\hat{X}H_{c}(y^{0})=H_{c}(y^{0})-\int d^{D-1}y\chi _{i}^{\left( 1\right)
}\left( y\right) \left[ G^{i}\left( y\right) ,H_{c}(y^{0})\right]  \notag \\
&&+\frac{1}{2}\int d^{D-1}yd^{D-1}z\chi _{i}^{\left( 1\right) }\left(
y\right) \chi _{j}^{\left( 1\right) }\left( y^{0},\mathbf{z}\right) \left[
G^{i}\left( y\right) \left[ G^{j}\left( y^{0},\mathbf{z}\right) ,H_{c}y^{0}%
\right] \right] -\cdots  \notag \\
&=&H_{c}(y^{0})-\int d^{D-1}y\left[ \pi _{0i}\partial _{j}A^{ji}-\frac{1}{%
2m^{2}}\partial _{i}\pi _{j0}\partial ^{\lbrack i}\pi ^{j]0}\right] .
\label{hamclsI}
\end{eqnarray}%
Clearly, the first-class constraint set (\ref{cein}) is irreducible (all the
equations are independent). This ends the GU procedure. In the sequel we
will improve it by passing to another first-class system (equivalent with
the original, second-class one at both classical and path integral levels)
such that the corresponding path integral takes a manifestly Lorentz
covariant form.

It is well known that any irreducible set of constraints can always be
replaced by a reducible one by introducing constraints that are consequences
of the ones already at hand (see Ref.~\cite{marc}, Chapter 1, Subsection 1.1.8).
In view of this, we supplement (\ref{cein}) with one more constraint, $%
G\equiv -m^{2}\partial _{i}G^{i}\approx 0$, such that the new constraint set%
\begin{eqnarray}
G^{i} &\equiv &-\frac{1}{m^{2}}\left( 2\partial _{j}\pi
^{ji}-m^{2}A^{0i}\right) \approx 0,  \label{cered1} \\
G &\equiv &-m^{2}\partial _{i}A^{0i}\approx 0  \label{cered2}
\end{eqnarray}%
remains first-class and, moreover, becomes off-shell first-order reducible.
This means that there exists a single relation among the constraint
functions involved in (\ref{cered1}) and (\ref{cered2}) which is strongly
equal to zero. In other words, if we organize the constraint functions (\ref%
{cered1}) and respectively (\ref{cered2}) into a column vector $G^{\kappa }$%
, then there exists a row vector $Z_{\kappa }$ (first-order reducibility
functions) such that $Z_{\kappa }G^{\kappa }=0$ in condensed De Witt
notations. Indeed, it is simple to check that one can choose%
\begin{equation}
\left( Z_{\kappa }\right) =\left(
\begin{array}{cc}
\partial _{i} & \frac{1}{m^{2}}%
\end{array}%
\right) .  \label{relred}
\end{equation}%
Obviously, (\ref{hamclsI}) is still a first-class Hamiltonian with respect
to the reducible first-class constraint set (\ref{cered1}) and (\ref{cered2}%
). This procedure preserves the classical equivalence with the first-class
theory from the GU method since it merely adds to it a combination of
existing first-class constraints, so it does not change either the number of
physical degrees of freedom or the classical observables, and keeps the
first-class Hamiltonian, such that the evolution is not affected. As a
result, the GU and first-order reducible first-class systems remain
equivalent also at the level of the Hamiltonian path integral quantization.
This further implies, given the established equivalence between the GU
first-class system and the original second-class theory, that the
first-order reducible first-class system is completely equivalent with the
original second-class theory.

At this stage, it is useful to make the canonical transformation%
\begin{equation}
A_{0i}\longrightarrow -\frac{1}{m^{2}}\Pi _{i},\qquad \pi
^{0i}\longrightarrow m^{2}B^{i},  \label{transf1}
\end{equation}%
which induces the non-vanishing Poisson brackets%
\begin{equation}
\left[ B^{i}(x),\Pi _{j}(y)\right] _{x^{0}=y^{0}}=\delta _{j}^{i}\delta
\left( \mathbf{x}-\mathbf{y}\right) .  \label{PBPiB}
\end{equation}%
It is important to remark that canonical transformations do not change
either the first-class behaviour or the reducibility. Consequently, the
constraints (\ref{cered1}) and (\ref{cered2}) become%
\begin{eqnarray}
G^{i} &\equiv &-\frac{1}{m^{2}}\left( 2\partial _{j}\pi ^{ji}+\Pi
^{i}\right) \approx 0,  \label{cered1n} \\
G &\equiv &\partial _{i}\Pi ^{i}\approx 0  \label{cered2n}
\end{eqnarray}%
and remain first-class, while the first-class Hamiltonian (\ref{hamclsI})
takes the form
\begin{eqnarray}
H_{GU}(y^{0}) &=&\int d^{D-1}y\left[ -\pi _{ij}\pi ^{ij}+\frac{1}{12}%
F_{ijk}F^{ijk}+\frac{m^{2}}{4}A_{ij}A^{ij}+\frac{m^{2}}{2}A_{ij}\partial
^{\lbrack i}B^{j]}\right.  \notag \\
&&\left. +\frac{m^{2}}{4}\partial _{\lbrack i}B_{j]}\partial ^{\lbrack
i}B^{j]}-\frac{1}{2m^{2}}\Pi _{i}\Pi ^{i}+\frac{1}{m^{2}}\Pi _{i}\left(
2\partial _{j}\pi ^{ji}+\Pi ^{i}\right) \right]  \label{hamclsIn}
\end{eqnarray}%
and is of course a first-class Hamiltonian with respect to (\ref{cered1n})
and (\ref{cered2n}). In addition, (\ref{relred}) remain first-order
reducibility functions for the constraint set (\ref{cered1n}) and (\ref%
{cered2n}).

Due to the equivalence between the first-order reducible first-class system
and the original second-class theory argued previously, one can replace the
Hamiltonian path integral of massive $2$-forms with that associated
with the reducible first-class system. The first-class Hamiltonian (\ref%
{hamclsIn}) outputs the argument of the exponential from the Hamiltonian
path integral of the reducible first-class system as%
\begin{eqnarray}
S_{GU} &=&\int d^{D}x\left[ \left( \partial _{0}A_{ij}\right) \pi
^{ij}+\left( \partial _{0}B_{i}\right) \Pi ^{i}+\pi _{ij}\pi ^{ij}-\frac{1}{%
12}F_{ijk}F^{ijk}-\frac{m^{2}}{4}A_{ij}A^{ij}\right.  \notag \\
&&-\frac{m^{2}}{2}A_{ij}\partial ^{\lbrack i}B^{j]}-\frac{m^{2}}{4}\partial
_{\lbrack i}B_{j]}\partial ^{\lbrack i}B^{j]}+\frac{1}{2m^{2}}\Pi _{i}\Pi
^{i}-\frac{1}{m^{2}}\Pi _{i}\left( 2\partial _{j}\pi ^{ji}+\Pi ^{i}\right)
\notag \\
&&\left. +\frac{1}{m^{2}}\lambda _{i}\left( 2\partial _{j}\pi ^{ji}+\Pi
^{i}\right) -\lambda \left( \partial _{i}\Pi ^{i}\right) \right] ,
\label{acttotint}
\end{eqnarray}%
where $\lambda _{i}$ and $\lambda $ denote the Lagrange multipliers
respectively corresponding to the first-class constraints (\ref{cered1n})
and (\ref{cered2n}). If we perform the transformation
\begin{equation}
\Pi ^{i}\longrightarrow \Pi ^{i},\qquad \lambda _{i}\longrightarrow \bar{%
\lambda}_{i}=\lambda _{i}-\Pi _{i}  \label{transf2}
\end{equation}%
in the path integral, the argument of the exponential becomes%
\begin{eqnarray}
S_{GU}^{\prime } &=&\int d^{D}x\left[ \left( \partial _{0}A_{ij}\right) \pi
^{ij}+\left( \partial _{0}B_{i}\right) \Pi ^{i}+\pi _{ij}\pi ^{ij}-\frac{1}{%
12}F_{ijk}F^{ijk}-\frac{m^{2}}{4}A_{ij}A^{ij}\right.  \notag \\
&&-\frac{m^{2}}{2}A_{ij}\partial ^{\lbrack i}B^{j]}-\frac{m^{2}}{4}\partial
_{\lbrack i}B_{j]}\partial ^{\lbrack i}B^{j]}+\frac{1}{2m^{2}}\Pi _{i}\Pi
^{i}  \notag \\
&&\left. +\frac{1}{m^{2}}\bar{\lambda}_{i}\left( 2\partial _{j}\pi ^{ji}+\Pi
^{i}\right) -\lambda \left( \partial _{i}\Pi ^{i}\right) \right] .
\label{acttotint1}
\end{eqnarray}%
At this stage, the reducible first-class system is endowed with the
Hamiltonian path integral
\begin{equation}
Z_{GU}=\int \mathcal{D}\left( fields \right) \mu \left( [A_{ij}], [B_{i}]\right) \exp
\left( \mathrm{i}S_{GU}^{\prime }\right) ,  \label{piGU}
\end{equation}%
where by `$fields$' we denoted the present fields, the associated momenta
and the Lagrange multipliers, and by `$\mu \left( [A_{ij}], [B_{i}]\right) $' the
integration measure associated with the model subject to the reducible
first-class constraints (\ref{cered1}) and (\ref{cered2}). This measure
includes some suitable canonical gauge conditions, is independent of gauge-fixing conditions \cite{FradkinVilk} and is chosen such that (%
\ref{piGU}) is convergent \cite{marcel}. A set of canonical gauge conditions associated to the first-class constraints (\ref{cered1}) and (\ref{cered2}) reads as
\begin{eqnarray}
\bar{C}_{i} &\equiv & \partial^{j}A_{ji}+B_{i} \approx 0,  \label{gf1} \\
\bar{C} &\equiv &\partial^{i}B_{i}\approx 0.  \label{gf2}
\end{eqnarray}%

In order to infer from (\ref{piGU}) a path integral that leads, after
integrating out the auxiliary variables, a manifestly Lorentz covariant
functional in its exponential, we enlarge the original phase-space with the
Lagrange multipliers $\bar{\lambda}_{i}$ and $\lambda $ respectively
associated with the first-class constraints (\ref{cered1}) and (\ref{cered2}%
) \cite{marc} (Chapter 11, Subsection 11.3.2) and with their canonical
momenta $p^{i}$ and $p$. We add the constraints
\begin{equation}
p^{i}\approx 0,\qquad p\approx 0,  \label{constrn}
\end{equation}%
such that the constraint set (\ref{cered1}), (\ref{cered2}), and (\ref%
{constrn}) is again first-class and off-shell first-order reducible. Adding
the supplementary first-class constraints (\ref{constrn}) does not alter the
established equivalence with the original second-class theory. Consequently,
the argument of the exponential from the Hamiltonian path integral for the
first-class theory with the phase-space locally parameterized by the
fields/momenta $\left\{ A_{ij},B_{i},\bar{\lambda}_{i},\lambda ,\pi
^{ij},\Pi ^{i},p^{i},p\right\} $ and subject to the first-class constraints (%
\ref{cered1n}), (\ref{cered2n}), and (\ref{constrn}) reads as%
\begin{eqnarray}
S_{GU}^{\prime \prime } &=&\int d^{D}x\left[ \left( \partial
_{0}A_{ij}\right) \pi ^{ij}+\left( \partial _{0}B_{i}\right) \Pi ^{i}+\left(
\partial _{0}\bar{\lambda}_{i}\right) p^{i}+\left( \partial _{0}\bar{\lambda}%
\right) p+\pi _{ij}\pi ^{ij}\right.  \notag \\
&&-\frac{1}{12}F_{ijk}F^{ijk}-\frac{m^{2}}{4}A_{ij}A^{ij}-\frac{m^{2}}{2}%
A_{ij}\partial ^{\lbrack i}B^{j]}-\frac{m^{2}}{4}\partial _{\lbrack
i}B_{j]}\partial ^{\lbrack i}B^{j]}  \notag \\
&&\left. +\frac{1}{2m^{2}}\Pi _{i}\Pi ^{i}+\frac{1}{m^{2}}\bar{\lambda}%
_{i}\left( 2\partial _{j}\pi ^{ji}+\Pi ^{i}\right) -\lambda \left( \partial
_{i}\Pi ^{i}\right) -\Lambda _{i}p^{i}-\Lambda p\right] .  \label{acttot}
\end{eqnarray}%
Performing in (\ref{acttot}) the integration over $\pi ^{ij}$, $\Pi ^{i}$, $%
p^{i}$, $p$, $\Lambda _{i}$, and $\Lambda $, the argument of the exponential
from the Hamiltonian path integral becomes
\begin{eqnarray}
S_{GU}^{\prime \prime \prime } &=&\int d^{D}x\left[ -\frac{1}{12}%
F_{ijk}F^{ijk}-\frac{m^{2}}{4}A_{ij}A^{ij}-\frac{m^{2}}{2}A_{ij}\partial
^{\lbrack i}B^{j]}-\frac{m^{2}}{4}\partial _{\lbrack i}B_{j]}\partial
^{\lbrack i}B^{j]}\right.  \notag \\
&&-\frac{1}{4}\left( \partial _{0}A_{ij}-\frac{1}{m^{2}}\partial _{\lbrack i}%
\bar{\lambda}_{j]}\right) \left( \partial ^{0}A^{ij}-\frac{1}{m^{2}}\partial
^{\lbrack i}\bar{\lambda}^{j]}\right) -\bar{\lambda}_{i}\left( \partial
^{0}B^{i}+\partial ^{i}\lambda \right)  \notag \\
&&\left. -\frac{1}{2m^{2}}\bar{\lambda}_{i}\bar{\lambda}^{i}-\frac{m^{2}}{2}%
\left( \partial _{0}B_{i}+\partial _{i}\lambda \right) \left( \partial
^{0}B^{i}+\partial ^{i}\lambda \right) \right] .  \label{acttot1}
\end{eqnarray}%
If we make the notations
\begin{equation}
\frac{1}{m^{2}}\bar{\lambda}_{i}\equiv -\bar{A}_{i0},\qquad \lambda \equiv
-B_{0},  \label{not}
\end{equation}%
then (\ref{acttot1}) can be written as%
\begin{eqnarray}
S_{GU}^{\prime \prime \prime } &=&\int d^{D}x\left[ -\frac{1}{12}%
F_{ijk}F^{ijk}-\frac{m^{2}}{4}A_{ij}A^{ij}-\frac{m^{2}}{2}A_{ij}\partial
^{\lbrack i}B^{j]}-\frac{m^{2}}{4}\partial _{\lbrack i}B_{j]}\partial
^{\lbrack i}B^{j]}\right.  \notag \\
&&-\frac{1}{4}\left( \partial _{0}A_{ij}+\partial _{\lbrack i}\bar{A}%
_{j]0}\right) \left( \partial ^{0}A^{ij}+\partial ^{\lbrack i}\bar{A}%
^{j]0}\right) -m^{2}\bar{A}_{i0}\left( \partial ^{i}B^{0}-\partial
^{0}B^{i}\right)  \notag \\
&&\left. -\frac{m^{2}}{2}\bar{A}_{i0}\bar{A}^{i0}-\frac{m^{2}}{2}\left(
\partial _{0}B_{i}-\partial _{i}B_{0}\right) \left( \partial
^{0}B^{i}-\partial ^{i}B^{0}\right) \right] ,  \label{acttot2}
\end{eqnarray}%
or, equivalently, as
\begin{eqnarray}
S_{GU}^{\prime \prime \prime } &=&\int d^{D}x\left[ -\frac{1}{12}%
F_{ijk}F^{ijk}-\frac{1}{4}\bar{F}_{0ij}\bar{F}^{0ij}-\frac{m^{2}}{4}%
A_{ij}A^{ij}-\frac{m^{2}}{2}\bar{A}_{i0}\bar{A}^{i0}\right.  \notag \\
&&\left. -\frac{m^{2}}{2}A_{ij}F^{ij}-m^{2}\bar{A}_{i0}F^{i0}-\frac{m^{2}}{4}%
F_{ij}F^{ij}-\frac{m^{2}}{2}F_{0i}F^{0i}\right] ,  \label{acttot3}
\end{eqnarray}%
where
\begin{eqnarray}
\bar{F}_{0ij} &=&\partial _{0}A_{ij}+\partial _{\lbrack i}\bar{A}_{j]0},
\label{altenot1} \\
F_{ij} &=&-\frac{1}{m}\partial _{\lbrack i}B_{j]},\qquad F_{0i}=-\frac{1}{m}%
\left( \partial _{0}B_{i}-\partial _{i}B_{0}\right) .  \label{altenot2}
\end{eqnarray}%
The functional (\ref{acttot3}) associated with the reducible first-class
system takes now a manifestly Lorentz covariant form%
\begin{equation}
S_{GU}^{\prime \prime \prime }\left[ \bar{B}_{\mu },\bar{A}_{\mu \nu }\right]
=\int d^{D}x\left[ -\frac{1}{12}\bar{F}_{\mu \nu \rho }\bar{F}^{\mu \nu \rho
}-\frac{1}{4}\left( F_{\mu \nu }-m\bar{A}_{\mu \nu }\right) \left( F^{\mu
\nu }-m\bar{A}^{\mu \nu }\right) \right] ,  \label{acttot4}
\end{equation}%
with%
\begin{gather}
\bar{A}_{\mu \nu }=-\bar{A}_{\nu \mu },\qquad \bar{A}_{\mu \nu }\equiv
\left( \bar{A}_{0j},A_{jk}\right) ,\qquad \bar{F}_{\mu \nu \rho }=\partial
_{\lbrack \mu }\bar{A}_{\nu \rho ]},  \label{explabarf} \\
\bar{B}_{\mu }=-\frac{1}{m}B_{\mu },\qquad F_{\mu \nu }=\partial _{\lbrack
\mu }\bar{B}_{\nu ]},  \label{explbbar}
\end{gather}%
and describes precisely the (Lagrangian) St\"{u}ckelberg coupling \cite%
{stueck} between the one-form $\bar{B}_{\mu }$ and the two-form $\bar{A}%
_{\mu \nu }$.

\subsection{BF method}

In the sequel we apply the BF method exposed in subsection \ref{BFmeth} to
massive $2$-forms. In view of this, we enlarge the original
phase-space by adding the bosonic fields/momenta $\left( B^{\mu },\Pi _{\mu
}\right) _{\mu =\overline{0,D-1}}$, endowed with the non-vanishing Poisson
brackets%
\begin{equation}
\left[ B^{\mu }(x),\Pi _{\nu }(y)\right] _{x^{0}=y^{0}}=\delta _{\nu }^{\mu
}\delta \left( \mathbf{x}-\mathbf{y}\right) .  \label{bf17}
\end{equation}%
The constraints (\ref{bf14}) gain in this case the concrete form $%
G_{A}\equiv \left( G^{(1)j},G_{j}^{(2)},G\right) \approx 0$, where%
\begin{eqnarray}
G^{(1)j} &\equiv &\chi ^{(1)j}+mB^{j}\approx 0,  \label{bf18a} \\
G_{j}^{(2)} &\equiv &\chi _{j}^{(2)}-m\Pi _{j}\approx 0,  \label{bf18b} \\
G &\equiv &\Pi _{0}\approx 0.  \label{bf18c}
\end{eqnarray}%
It is easy to check that they form an Abelian and irreducible first-class
constraint set. The first-class Hamiltonian complying with the general
requirements (\ref{bf12}) and (\ref{bf13}) is expressed by%
\begin{eqnarray}
H_{BF}(x^{0}) &=&H_{c}(x^{0})+\int d^{D-1}x\left[ \frac{1}{2}\Pi ^{i}\Pi
_{i}-\frac{1}{m}\Pi ^{j}\chi _{j}^{(2)}\right.  \notag \\
&&\left. -\frac{1}{2}\left( mA^{jk}-\frac{1}{2}\partial ^{\lbrack
j}B^{k]}\right) \partial _{\lbrack j}B_{k]}-B^{0}\partial ^{j}\left( \Pi
_{j}+mA_{0j}\right) \right] .  \label{bf19}
\end{eqnarray}%
Consequently, the Hamiltonian gauge algebra relations (\ref{bf13}) are given
by
\begin{eqnarray}
\left[ H_{BF}(x^{0}),G^{(1)j}(x)\right] &=&0=\left[
H_{BF}(x^{0}),G_{j}^{(2)}(x)\right] ,  \label{bf20} \\
\left[ H_{BF}(x^{0}),G(x)\right] &=&\frac{1}{m}\partial ^{j}G_{j}^{(2)}(x).
\label{bf21}
\end{eqnarray}

In the following we analyze the Hamiltonian path integral for the above BF
first-class system, equivalent with that of massive $2$-forms.
Imposing some appropriate gauge-fixing conditions $C_{A}\equiv \left(
C_{j}^{(1)},C^{(2)j},C\right) \approx 0$, the Hamiltonian path integral
takes the form%
\begin{equation}
Z_{BF}=\int \mathcal{D}\left( A^{\mu \nu },\pi _{\mu \nu },B^{\mu },\Pi
_{\mu }\right) \left( \prod\limits_{A,B}\delta \left( G_{A}\right) \delta
\left( C_{B}\right) \right) \left( \det \left( \left[ G_{A^{\prime
}},C_{B^{\prime }}\right] \right) \right) \exp \left( \mathrm{i}%
S_{BF}\right) ,  \label{bf22}
\end{equation}%
where%
\begin{equation}
S_{BF}=\int d^{D}x\left( \pi _{0j}\dot{A}^{0j}+\pi _{jk}\dot{A}^{jk}+\Pi
_{\mu }\dot{B}^{\mu }-H_{BF}\right) .  \label{bf23}
\end{equation}%
In this situation an example of canonical gauge conditions is
\begin{equation}
C_{j}^{(1)} \equiv \Pi _{j}\approx 0, \qquad C^{(2)j} \equiv B^{j}\approx 0, \qquad G \equiv B^{0}\approx 0.  \label{GFn}
\end{equation}
By performing a Fourier representation of the factors $\delta \left(
G_{A}\right) $ from (\ref{bf22}), it becomes%
\begin{eqnarray}
Z_{BF} &=&\int \mathcal{D}\left( A^{\mu \nu },\pi _{\mu \nu },B^{\mu },\Pi
_{\mu },\lambda ^{(1)j},\lambda ^{(2)j},\lambda \right) \left(
\prod\limits_{A}\delta \left( C_{A}\right) \right) \times  \notag \\
&&\times \left( \det \left( \left[ G_{A^{\prime }},C_{B^{\prime }}\right]
\right) \right) \exp \left( \mathrm{i}S_{BF}^{\prime }\right) ,  \label{bf24}
\end{eqnarray}%
with%
\begin{equation}
S_{BF}^{\prime }=\int d^{D}x\left( \pi _{0j}\dot{A}^{0j}+\pi _{jk}\dot{A}%
^{jk}+\Pi _{\mu }\dot{B}^{\mu }-H_{BF}-\sum\limits_{m=1}^{2}\lambda
^{(m)j}G_{j}^{(m)}-\lambda G\right) .  \label{bf25}
\end{equation}%
Employing in (\ref{bf24}) the change of variables%
\begin{equation}
\Pi _{j}\longrightarrow \Pi _{j}^{\prime }=\Pi _{j}+mA_{0j},\qquad \pi
_{0j}\longrightarrow \pi _{0j}^{\prime }=\pi _{0j}+mB_{j}  \label{bf26}
\end{equation}%
and integrating over the momenta $\pi _{jk}$, the argument of the
exponential from the path integral reads as%
\begin{eqnarray}
S_{BF}^{\prime \prime } &=&\int d^{D}x\left[ \pi _{0j}^{\prime }\dot{A}%
^{0j}+\Pi _{j}^{\prime }\dot{B}^{j}+\Pi _{0}\dot{B}^{0}-\frac{1}{12}%
F_{ijk}F^{ijk}-\frac{1}{2}\Pi _{j}^{\prime }\Pi ^{\prime j}\right.  \notag \\
&&-\frac{1}{4}\left( \dot{A}_{ij}-\frac{1}{m}\partial _{\lbrack i}\Pi
_{j]}^{\prime }-\partial _{\lbrack i}\lambda _{j]}^{(2)}\right) \left( \dot{A%
}^{ij}-\frac{1}{m}\partial ^{\lbrack i}\Pi ^{\prime j]}-\partial ^{\lbrack
i}\lambda ^{(2)j]}\right)  \notag \\
&&-\frac{1}{4}\left( \partial _{\lbrack i}B_{j]}-mA_{ij}\right) \left(
\partial ^{\lbrack i}B^{j]}-mA^{ij}\right) +B^{0}\partial ^{j}\Pi
_{j}^{\prime }  \notag \\
&&\left. -\lambda ^{(1)j}\pi _{0j}^{\prime }-m\lambda ^{(2)j}\Pi
_{j}^{\prime }-\lambda \Pi _{0}\right] .  \label{bf27}
\end{eqnarray}%
Making in the last form of the path integral the change of variables
\begin{equation}
\Pi _{j}^{\prime }\longrightarrow \bar{A}_{0j}\equiv \frac{1}{m}\Pi
_{j}^{\prime }+\lambda _{j}^{(2)},\qquad \lambda _{j}^{(2)}\longrightarrow
\lambda _{j}^{(2)}  \label{bf28}
\end{equation}%
and using the notations (\ref{explabarf}), the argument of the exponential
from the path integral is turned into%
\begin{eqnarray}
S_{BF}^{\prime \prime \prime } &=&\int d^{D}x\left[ \pi _{0j}^{\prime }\dot{A%
}^{0j}+m\left( \bar{A}_{0j}-\lambda _{j}^{(2)}\right) \dot{B}^{j}+\Pi _{0}%
\dot{B}^{0}-\frac{1}{12}\bar{F}_{\mu \nu \rho }\bar{F}^{\mu \nu \rho }\right.
\notag \\
&&-\frac{1}{4}\left( \partial _{\lbrack i}B_{j]}-mA_{ij}\right) \left(
\partial ^{\lbrack i}B^{j]}-mA^{ij}\right) -\frac{m^{2}}{2}\left( \bar{A}%
_{0j}\bar{A}^{0j}-\lambda _{j}^{(2)}\lambda ^{(2)j}\right)  \notag \\
&&\left. -m\left( \bar{A}_{0j}-\lambda _{j}^{(2)}\right) \partial
^{j}B^{0}-\lambda ^{(1)j}\pi _{0j}^{\prime }-\lambda \Pi _{0}\right] .
\label{bf29}
\end{eqnarray}%
Finally, we integrate in the path integral over $\lambda ^{(2)j}$, $\pi
_{0j}^{\prime }$, $\lambda ^{(1)j}$, $\Pi _{0}$, $\lambda $, and $A^{0j}$,
such that the argument of the exponential reduces to
\begin{equation}
\tilde{S}_{BF}\left[ B_{\mu },\bar{A}_{\mu \nu }\right] =\int d^{D}x\left[ -%
\frac{1}{12}\bar{F}_{\mu \nu \rho }\bar{F}^{\mu \nu \rho }-\frac{1}{4}\left(
F_{\mu \nu }-m\bar{A}_{\mu \nu }\right) \left( F^{\mu \nu }-m\bar{A}^{\mu
\nu }\right) \right] ,  \label{bf30}
\end{equation}%
where $F_{\mu \nu }=\partial _{\lbrack \mu }B_{\nu ]}$. It is now obvious
that the path integral of the BF first-class system takes a manifestly
Lorentz covariant form and describes again the (Lagrangian) St\"{u}ckelberg
coupling between the one-form $B_{\mu }$ and the two-form $\bar{A}_{\mu \nu
} $.

\section{Conclusion\label{sect4}}

In this paper we analyzed massive $2$-form fields from the
point of view of gauge-unfixing and respectively Batalin--Fradkin methods.
The first approach (GU) relies on separating the (independent) second-class
constraints into two subsets, of which one is first-class and the other a
set of canonical gauge conditions. Starting from the original canonical
Hamiltonian, we generated a first-class Hamiltonian with respect to the
first-class constraint subset. Finally, we built the Hamiltonian path
integral of the GU first-class system and then eliminated the auxiliary
fields and performed some variable redefinitions such that the path integral
finally takes a manifestly Lorentz covariant form. The second approach (BF)
involves an appropriate extension of the original phase-space and then the
construction of a first-class system on the extended phase-space that
reduces to the original, second-class theory in the zero limit of all
extravariables. The Hamiltonian path integral of the BF first-class system
leads again, after integrating out some of the variables and performing some
field redefinitions, to a manifestly Lorentz covariant form. It is
interesting to remark that both approaches require an appropriate extension
of the phase-space in order to render a manifestly covariant path integral.
Both procedures allowed the identification of the Lagrangian path integral
for St\"{u}ckelberg-coupled $1$- and $2$-forms.

\section*{Acknowledgement}

The authors wish to thank Constantin Bizdadea and Odile Saliu for useful discussions
and comments. Two authors (E. M. C. and S. C. S.) acknowledge partial support from the
contract 2-CEx-06-11-92/2006 with the Romanian Ministry of Education and
Research (M.Ed.C.).

\end{document}